\documentclass[twocolumn,superscriptaddress,amsmath,amssymb,pra,longbibliography]{revtex4-1}
\usepackage{amssymb,amsmath}
\DeclareMathAlphabet\mathbfcal{OMS}{cmsy}{b}{n}
\usepackage{cmap}
\usepackage{graphicx} 
\usepackage{bm}% bold math
\usepackage{color}
\usepackage{blindtext}
\usepackage{hyperref}
\usepackage{listings}
\usepackage{booktabs}
\usepackage{multirow}
\usepackage{amsmath}
\usepackage{mathrsfs}
\usepackage{braket}
\usepackage{mathtools}

\usepackage{dsfont}
\usepackage{tabularx}

\newcommand{\blue}[1]{{\color{black}{#1}}}

\begin{document}

%\title{The enigma of orbital angular momentum of spatiotemporal vortex pulses}

\title{Orbital angular momentum of optical, acoustic, and quantum-mechanical spatiotemporal vortex pulses}

\author{Konstantin Y. Bliokh}
\affiliation{Theoretical Quantum Physics Laboratory, RIKEN Cluster for Pioneering Research, Wako-shi, Saitama 351-0198, Japan}
\affiliation{Donostia International Physics Center (DIPC), Donostia-San Sebasti\'{a}n 20018, Spain}
%\author{Howard M. Milchberg}
%\affiliation{Institute for Research in Electronics and Applied Physics, University of Maryland, College Park, Maryland 20742, USA}
%\author{Franco Nori}
%\affiliation{Theoretical Quantum Physics Laboratory, RIKEN Cluster for Pioneering Research, Wako-shi, Saitama 351-0198, Japan}
%\affiliation{Physics Department, University of Michigan, Ann Arbor, Michigan 48109-1040, USA}

%\date{\today}

\begin{abstract}
Motivated by recent progress in the generation of optical spatiotemporal vortex pulses (STVPs), there is a theoretical discussion about the transverse orbital angular momentum (OAM) carried by such pulses. Two recent works [K. Y. Bliokh, Phys. Rev. Lett. {\bf 126}, 243601 (2021)] and [S. W. Hancock {\it et al.}, Phys. Rev. Lett. {\bf 127}, 193901 (2021)] claimed the OAM values which differ by a factor of 2 for circular STVPs. Here we resolve this controversy by showing that the result by Hancock {\it et al.} is correct for the {\it total} OAM, while the result by Bliokh describes the suitably defined {\it intrinsic} part of the OAM. The other, {\it extrinsic} part of the OAM originates from the fact that plane waves of the same amplitude but different frequencies in the pulse spectrum contain different densities of photons, which induces a transverse vortex-dependent shift of the photon centroid even in a STVP with symmetric energy-density distribution. We describe similar peculiarities of acoustic and quantum-relativistic (Klein-Gordon and arbitrary-spin) STVPs. In all cases, only the intrinsic OAM keeps a universal form independent of the details of the problem and similar to the OAM of monochromatic vortex beams.
\end{abstract}

%\keywords{Acoustic force; canonical momentum; acoustic spin; acoustic torque.}

\maketitle

%%%%%%%%%%%%%%%%%%%%%%%%%%%%%%
\section{Introduction}
%%%%%%%%%%%%%%%%%%%%%%%%%%%%%%
Monochromatic vortex beams with screw-type phase singularities (vortices) have been extensively explored in optics \cite{Allen1992, Allen_book, Bekshaev_book, Andrews_book, Molina2007, Franke2008, Bliokh2015PR}, acoustics \cite{Hefner1999, Lekner2006, Volke2008, Demore2012, Anhauser2012, Marzo2018,Bliokh2019_II}, and quantum physics \cite{Bliokh2007, Uchida2010, Verbeeck2010, McMorran2011, Bliokh2017, Lloyd2017, Larocque2018, Clark2015, Luski2021} in the past 30 years. Such beams with circularly-symmetric transverse intensity distributions carry an intrinsic orbital angular momentum (OAM) along the beam axis. In the paraxial regime, the OAM value is $\ell$ per particle (photon, phonon, electron, etc.) in the $\hbar=1$ units, where $\ell$ is the integer topological charge of the vortex. 
More recently, there was a theoretical \cite{Sukhorukov2005, Dror2011, Bliokh2012, Bliokh2021, Hancock2021} and experimental \cite{Jhajj2016, Chong2020, Hancock2019, Gui2021, Hancock2020, Fang2021, Wang2021, Zhang2022, Huang2022} progress in studies of {\it spatiotemporal vortex pulses} (STVPs) with edge-type (or, generally, mixed edge-screw) phase dislocations, i.e., vortices in the space-time domain. Such pulses carry the OAM orthogonal to the pulse plane. STVPs can be considered in a wider context of structured space-time waves \cite{Yessenov2022}.

It is natural to assume that the OAM of STVPs with circular intensity distributions is $\ell$ per particle, and the first calculations indeed provided this value \cite{Bliokh2012,Chong2020,Bliokh2021}. However, recent calculations for optical STVPs \cite{Hancock2021} reported an unusual expression for the OAM corresponding to $\ell/2$ per photon in circular pulses. 

Here we re-examine the OAM of STVPs and find that the result of \cite{Hancock2021} is correct and describes the {\it total} OAM of an optical STVP with symmetric energy-density distribution. Yet, previous results \cite{Bliokh2012,Chong2020,Bliokh2021} are also valid for the {\it intrinsic} part of the OAM. The remaining {\it extrinsic} part originates from the transverse vortex-dependent shift of the `photon (probability) centroid'. This shift appears because plane waves of the same amplitude but different frequencies in the pulse spectrum contain different densities of photons \cite{CT,BB1996}, and positions of the energy-density and probability centroids differ in the general relativistic case \cite{Bliokh2012PRL,Smirnova2018}. 

We also examine this problem for acoustic and relativistic quantum-mechanical (Klein-Gordon and arbitrary-spin) STVPs. We show that the total and extrinsic OAM of spatiotemporal vortices are very sensitive to the spin, choice of the wavefunction, and the mean momentum of the pulse, whereas the intrinsic OAM, defined with respect to the particle probability centroid, is a robust universal quantity,  similar to the intrinsic OAM of monochromatic vortex beams. 

In contrast to phase singularities in monochromatic waves, spatiotemporal vortices do not have a unique well-defined position. For example, a single optical charge-$\ell$ vortex in the electric field generally splits into a constellation of $\ell$ charge-1 vortices in the vector-potential field. Therefore, it is {\it impossible} to construct a globally symmetric STVP with a uniquely defined center (e.g., the electric-field and vector-potential distributions inevitably differ from each other). Only the use of the particle probability centroid simplifies the OAM description in the general case.

%%%%%%%%%%%%%%%%%%%%%%%%%%%%%%%%%%%%%%
\begin{figure*}[t!]
\includegraphics[width=0.95\linewidth]{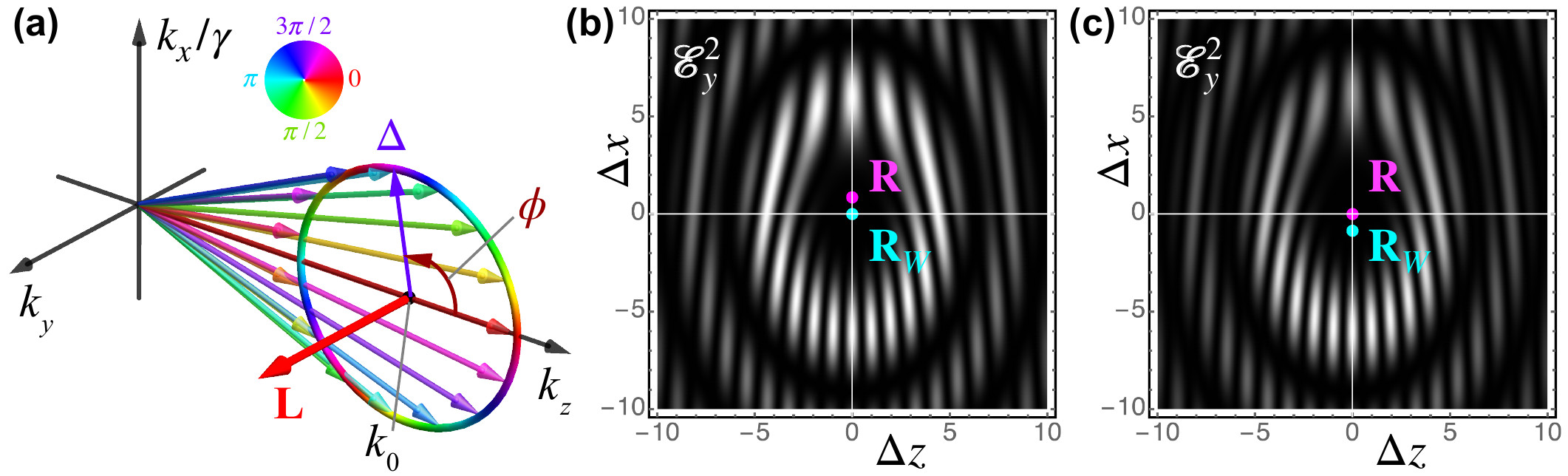}
\caption{
(a) The plane-wave spectrum of the linearly $y$-polarized Bessel-like STVP (\ref{eq_spectrum}); the phases of the plane-wave amplitudes $E_y  \propto \exp(i\ell\phi)$ are color-coded. (b) Its instantaneous ($t=0$) real-space electric-field intensity ${\mathcal E}_y^2({\bf r})$ (or the similar energy density ${\mathbfcal E}^2 + {\mathbfcal H}^2$) distribution with the positions of the `photon probability centroid' ${\bf R}$, Eqs.~(\ref{eq_R}) and (\ref{eq_X}), and the energy centroid ${\bf R}_W$. (c) Same as (b) but for the STVP with $E_y \propto \sqrt{\omega} \exp(i\ell\phi)$ providing for equal densities of photons in each of the plane waves in the spectrum. The parameters are: $\ell = 3$, $\Delta/k_0 = 0.4$ (this not-too-small value is for the better visualization of the centroid shift), and $\gamma = 0.7$.
\label{Fig1}}
\end{figure*}
%%%%%%%%%%%%%%%%%%%%%%%%%%%%%%%%%%%%%%

%%%%%%%%%%%%%%%%%%%%%%%%%%%%%%
\section{General electromagnetic equations}
%%%%%%%%%%%%%%%%%%%%%%%%%%%%%%
We first examine electromagnetic pulses constructed as free-space solutions of Maxwell's equations. 
The spin and orbital parts of the angular momentum of light are well defined and separably observable \cite{Allen1992,Allen_book,Bekshaev_book,Andrews_book,Molina2007, Franke2008, Bliokh2015PR, Enk1994_II, ONeil2002, Garces2003, Bliokh2010, BB2011, Barnett2016, Leader2016}, particularly in the paraxial approximation where the spin and OAM are determined, respectively, by the polarization and complex phase-amplitude structure of the field. Since we are interested in the OAM-related controversy and not in the spin-related phenomena, we consider paraxial linearly-polarized wavepackets, where one can neglect the spin and spin-orbit interaction effects \cite{Bliokh2015NP,Bliokh2021}.

Using the real-valued electric and magnetic fields ${\mathbfcal{E}}({\bf r},t)$ and ${\mathbfcal H}({\bf r},t)$, as well as the corresponding complex plane-wave Fourier amplitudes ${\bf E}({\bf k}) e^{-i\omega({\bf k}) t}$ and ${\bf H}({\bf k}) e^{-i\omega({\bf k}) t}$ satisfying ${\bf H} = {\omega}^{-1}{\bf k} \times {\bf E}$ and $\omega({\bf k}) = k$ in the $c=1$ units, the normalized \blue{integral (expectation) values of the} energy, momentum, and (orbital) angular momentum of the wavepacket can be written as \cite{Jackson,CT}:
\begin{align}
\label{eq_W}
& W =  \frac{1}{2N} \int \!\left({\mathbfcal{E}}^2 + {\mathbfcal{H}}^2 \right) d^3{\bf r}
= \frac{\int |{\bf E}|^2 \, d^3{\bf k}}{\int \omega^{-1}|{\bf E}|^2 \, d^3{\bf k}} \simeq \omega_0\,,
\end{align}
\begin{align}
\label{eq_P}
&{\bf P} =  \frac{1}{N} \int ({\mathbfcal{E}} \times  {\mathbfcal{H}})\, d^3{\bf r} 
=  \frac{\int  \omega^{-1} {\bf k}\, |{\bf E}|^2 \, d^3{\bf k}}{\int \omega^{-1}|{\bf E}|^2 \, d^3{\bf k}}\simeq {\bf k}_0\,, \\
\label{eq_J}
&{\bf L} \simeq \frac{1}{N}\! \int {\bf r}\! \times \! ({\mathbfcal{E}} \times  {\mathbfcal{H}}) d^3{\bf r} 
\simeq \frac{ \int \! \omega^{-1} {\bf E}^*\! (- i {\bf k} \times \!{\bm \nabla}_{\bf k}) {\bf E} \, d^3{\bf k}}{\int \omega^{-1}|{\bf E}|^2 \, d^3{\bf k}}.
\end{align}
Here ${\bf k}_0$ and $\omega_0 = \omega ({\bf k}_0)$ are the central wave vector and frequency of the pulse, whereas $N \propto \int \omega^{-1} |{\bf E}|^2 \, d^3{\bf k}$ is the number of photons in the pulse.
Note also that the first expression in Eq.~(\ref{eq_J}) is the exact expression for the {\it total} angular momentum valid for arbitrary localized solution of Maxwell equations \cite{Jackson}, and the spin part is neglected in the transition to the second expression in Eq.~(\ref{eq_J}).
Equations (\ref{eq_W})--(\ref{eq_J}) represent the expectation values of the energy, $\omega$, momentum, $\hat{\bf p} = {\bf k}$, and OAM, $\hat{\bf L} = \hat{\bf r} \times \hat{\bf p} = - i ({\bf k} \times {\bm \nabla}_{\bf k})$, operators with the `photon wavefunction' ${\bf E}({\bf k})/\sqrt{\omega({\bf k})}$ in the momentum representation. The fact that the photon wavefunction is well-defined only in ${\bf k}$-space and with the factor $\omega^{-1}$ is well known \cite{BB1996,CT} and crucial for our problem. It means that the density of photons (well-defined in monochromatic fields) in a plane wave is $\rho \propto |{\bf E}|^2/\omega$. 

There is one more quantity important for our consideration: the `photon centroid'. Although the photon wavefunction and the corresponding probability density are generally ill-defined in real space (because the operator $1/{\omega({\bf k})}$ is {\it nonlocal} in ${\bf r}$-space) \cite{BB1996}, which is related to the Weinberg-Witten theorem \cite{Weinberg1980}, the {\it probability centroid} of the photon is well-defined as the normalized expectation value of the position operator $\hat{\bf r} = i {\bm \nabla}_{\bf k}$ \cite{Bliokh2010,Bliokh2012PRL,Smirnova2018}:
\begin{align}
\label{eq_R}
{\bf R} = \frac{ \int  \omega^{-1} {\bf E}^*e^{i\omega t} \cdot (i{\bm \nabla}_{\bf k}) {\bf E}e^{-i\omega t} \, d^3{\bf k}}{\int \omega^{-1} |{\bf E}|^2 \, d^3{\bf k}} \,.
\end{align}
Here we accounted for the time-dependent factor $e^{-i\omega t}$ in the field Fourier amplitudes, which provides for the propagation of the centroid ${\bf R}(t)$ with the mean value of the group velocity, $v_g({\bf k}) = {\bm \nabla}_{\bf k}\, \omega = {\bf k}/k$. This straightforward propagation is not important for our consideration below. Notably, the photon centroid (\ref{eq_R}) generally differs from the centroid of the energy density distribution, ${\bf R}_W$, defined similarly to Eq.~(\ref{eq_R}) but without the $\omega^{-1}$ factors in the integrands \cite{Bliokh2012PRL,Smirnova2018}. The photon centroid allows one to separate the intrinsic and extrinsic parts in the total OAM (\ref{eq_J}) \cite{Bliokh2010,Bliokh2012PRL,Bliokh2015PR,Smirnova2018}:
\begin{align}
\label{eq_ext}
{\bf L}^{\rm ext} = {\bf R} \times {\bf P}\,, \quad
{\bf L}^{\rm int} = {\bf L} - {\bf L}^{\rm ext} \,.
\end{align}
The change of the coordinate origin ${\bf r} \to {\bf r} + {\bf a}$ affects only the extrinsic OAM in agreement with t	he parallel-axis theorem: ${\bf L}^{\rm ext} \to  {\bf L}^{\rm ext} + {\bf a} \times {\bf P}$, ${\bf L}^{\rm int} \to {\bf L}^{\rm int}$. An alternative definition of the intrinsic OAM with respect to ${\bf R}_W$ is considered in the examples below.
 
%%%%%%%%%%%%%%%%%%%%%%%%%%%%%%
\section{OAM of optical STVPs}
%%%%%%%%%%%%%%%%%%%%%%%%%%%%%%
We are now in the position to analyze the OAM of an electromagnetic STVP. For simplicity, we consider a  Bessel-like $z$-propagating STVP \cite{Bliokh2021} with an elliptical \blue{intensity-distribution} shape in the $(z,x)$-plane, charge-$\ell$ vortex, $y$-directed OAM, and \blue{linear} $y$-polarization corresponding to the single electric-field component ${\cal E}_y$, Fig.~\ref{Fig1}. Its plane-wave spectrum lies on a ${\bf k}$-space ellipse and is described by:
\begin{equation}
\label{eq_spectrum}
k_z = k_0 + \Delta \cos \phi \,, ~~
k_x = \gamma \Delta \sin \phi \,, ~~
E_y \propto \exp(i\ell\phi)\,.
\end{equation}
Here $\Delta \ll k_0$ and $\gamma \Delta \ll k_0$ are the ellipse semiaxes along the $z$ and $x$ directions, respectively, and $\phi$ is the azimuthal angle with respect to the ellipse center, Fig.~\ref{Fig1}(a).

For the 1D elliptical spectrum (\ref{eq_spectrum}), all ${\bf k}$-space integrals (\ref{eq_W})--(\ref{eq_R}) are reduced to integrals over $\phi \in (0,2\pi)$. Substituting Eq.~(\ref{eq_spectrum}) with the relations 
\[ \dfrac{\partial}{\partial k_x}\! =\! \dfrac{\cos\phi}{\gamma \Delta} \dfrac{\partial}{\partial \phi}, ~ \dfrac{\partial}{\partial k_z} \!=\! - \dfrac{\sin\phi}{\Delta} \dfrac{\partial}{\partial \phi}, ~\dfrac{1}{\omega} \!\simeq \!\dfrac{1}{k_0} \!\left(1-\dfrac{\Delta \cos\phi}{ k_0}\right)\!,\] 
into Eq.~(\ref{eq_J}) we find the normalized OAM of the STVP:
\begin{align}
\label{eq_Ly}
{L}_y  
%= \frac{ \int_0^{2\pi} \frac{1}{\omega}\! \left[ e^{-i\ell\phi} \left( -i {\bf k} \times {\bm \nabla}_{\bf k}\right)_y e^{i\ell\phi} \right] d\phi}{\int \frac{1}{\omega}\, d\phi} 
\simeq  \frac{\gamma \ell}{2} \,.
\end{align}
This equation agrees with the calculations of \cite{Hancock2021} (using a completely different approach \blue{based on an OAM-like operator commuting with the effective $z$-evolution `Hamiltonian' of the paraxial wave equation}) and differs from all other calculations \cite{Bliokh2012,Chong2020,Bliokh2021}. In particular, for a circular pulse with $\gamma =1$ it yields {\it half-integer} OAM ${L}_y  = \ell/2$, in sharp contrast to the integer OAM of monochromatic vortex beams \cite{Allen1992, Allen_book, Bekshaev_book, Andrews_book, Molina2007, Franke2008, Bliokh2015PR, Hefner1999, Lekner2006, Volke2008, Demore2012, Anhauser2012, Marzo2018, Bliokh2019_II, Bliokh2007, Uchida2010, Verbeeck2010, McMorran2011, Bliokh2017, Lloyd2017, Larocque2018, Clark2015, Luski2021}. We checked numerically that the standard calculation of the total angular momentum of the pulse via real-space integrals in Eqs.~(\ref{eq_W}) and (\ref{eq_J}), i.e., $\omega_0 L_y / W$, yield the same result (\ref{eq_Ly}), and hence the spin-related contributions are negligible as expected.

To unveil the physical meaning of the unusual OAM (\ref{eq_Ly}), we calculate the photon centroid (\ref{eq_R}) in the STVP (\ref{eq_spectrum}). Surprisingly, it exhibits a {\it transverse vortex-dependent shift} along the $x$-axis, Fig.~\ref{Fig1}(b):
\begin{align}
\label{eq_X}
{X}
%= \frac{\int_0^{2\pi} \frac{1}{\omega}\! \left[ e^{-i\ell\phi} i \frac{\partial}{\partial k_x} e^{i\ell\phi} \right] d\phi}{\int \frac{1}{\omega}\, d\phi} 
\simeq \frac{\ell}{2\gamma k_0} \,.
\end{align}
This shift occurs even though the energy-density (or electric-field intensity $|{\mathbfcal E}|^2$) centroid is not shifted, ${X}_W =0$, because the STVP consists of multiple plane waves with equal electric-field amplitudes but slightly different frequencies and, hence, {\it different photon densities} $\rho \propto |{\bf E}|^2/\omega$. Interference of these waves, including the vortex phase factor, results in the shift (\ref{eq_X}). The vortex-induced transverse shifts of the probability centroid, different from the energy centroid, are typical for relativistic waves with OAM \cite{Bliokh2012PRL,Bliokh2012,Smirnova2018}.

The shift (\ref{eq_X}) together with the momentum of the pulse, ${P}_z \simeq k_0$, mean that the pulse carries {\it intrinsic} and {\it extrinsic} OAM (\ref{eq_ext}):
\begin{align}
\label{eq_Ly_ext}
{L}_y^{\rm int} \simeq  \ell\, \frac{\gamma + \gamma^{-1}}{2} \,, \quad
 {L}_y^{\rm ext} \simeq - k_0 {X} \simeq - \frac{\ell}{2\gamma} \,.
\end{align}
This intrinsic OAM is in precise agreement with calculations in \cite{Bliokh2012,Bliokh2021} and yields the usual integer value ${L}_y^{\rm int}  = \ell$ for circular STVPs with $\gamma =1$.

%%%%%%%%%%%%%%%%%%%%%%%%%%%%%%%%%%%%%%
\begin{figure}[t!]
\includegraphics[width=\linewidth]{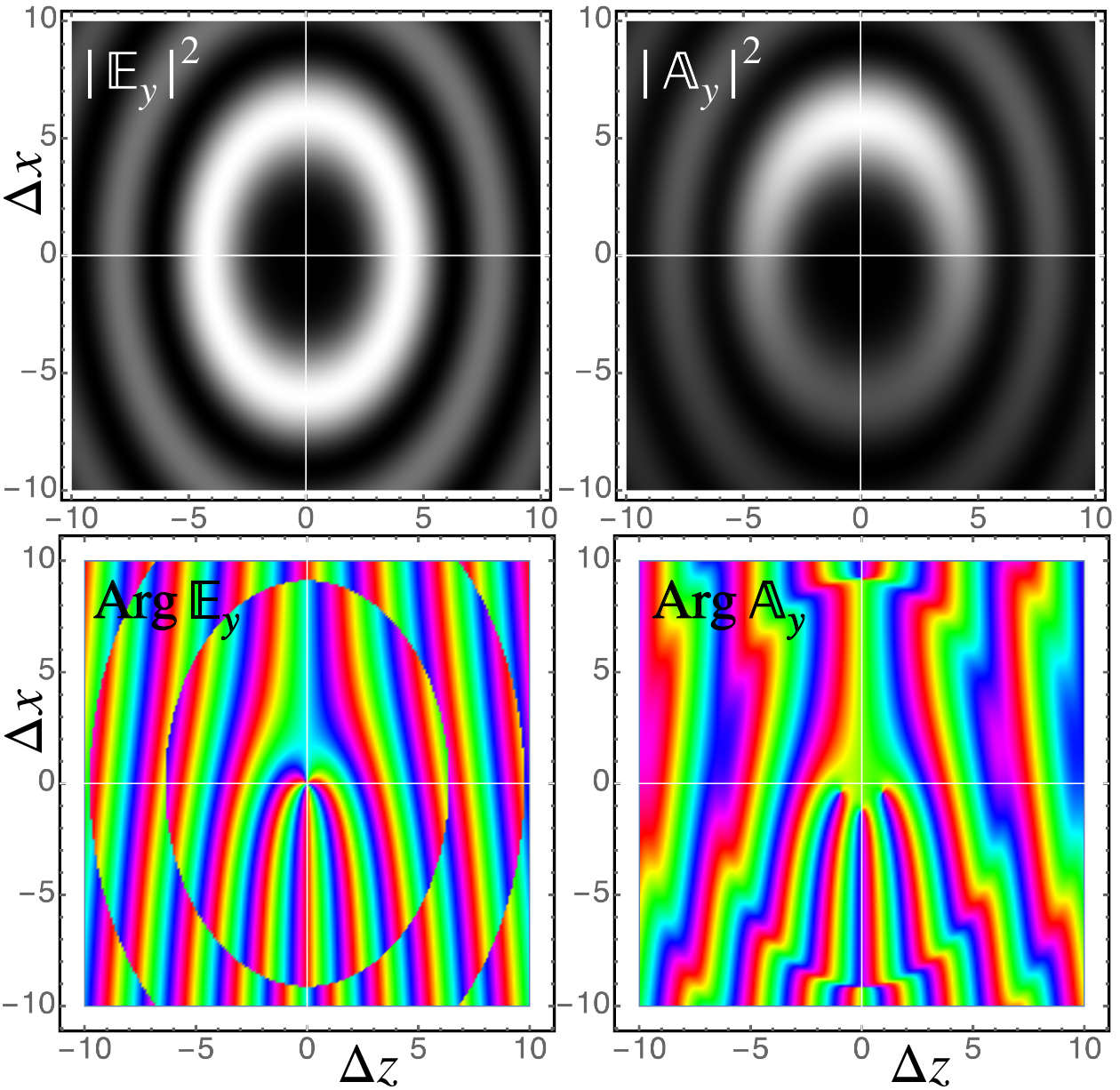}
\caption{
The instantaneous ($t=0$) real-space intensity and phase distributions of the complex electric field ${\mathbb E}_y({\bf r})$ (the complex Fourier transform of $E_y({\bf k})$; the real field in Fig.~\ref{Fig1}(b) is ${\mathcal E}_y = {\rm Re}\,{\mathbb E}_y$) and the corresponding Coulomb-gauge vector-potential ${\mathbb A}_y({\bf r})$ for the STVP (\ref{eq_spectrum}) shown in Fig.~\ref{Fig1}(b). Significant differences in these distributions reflect the impossibility of a globally symmetric spatiotemporal vortex with a uniquely defined singularity.
%the lack of global symmetry and uniquely defined phase singularity in spatiotemporal vortices.
\label{Fig2}}
\end{figure}
%%%%%%%%%%%%%%%%%%%%%%%%%%%%%%%%%%%%%%

Equations (\ref{eq_Ly})--(\ref{eq_Ly_ext}) resolve the controversy between \cite{Bliokh2012,Chong2020,Bliokh2021} and \cite{Hancock2021}. On the one hand, the total OAM is calculated correctly in \cite{Hancock2021}, while Refs.~\cite{Chong2020,Bliokh2021} did not properly account for the variations of frequency in the pulse spectrum. On the other hand, the intrinsic part of the OAM, defined via Eq.~(\ref{eq_ext}), is given by the expressions found in \cite{Bliokh2012,Bliokh2021} (Ref.~\cite{Chong2020} did not properly account for the ellipticity $\gamma$). 

There are several important remarks. First, one can construct a modified STVP with an additional amplitude factor $\sqrt{\omega}$ in the spectrum (\ref{eq_spectrum}): $E_y \propto \sqrt{\omega} \exp(i\ell\phi)$, so that all the interfering plane waves carry the same density of photons. This results in ${X} = {L}_y^{\rm ext} =0$ and ${L}_y = {L}_y^{\rm int}$ given by Eq.~(\ref{eq_Ly_ext}). Such a paraxial pulse looks very similar to the regular STVP, but with slight asymmetry of the energy-density distribution and shift of the {\it energy} centroid opposite to Eq.~(\ref{eq_X}): ${X}_W = - {\ell}/({2\gamma k_0})$, Fig.~\ref{Fig1}(c).

Second, for symmetric $z$-propagating monochromatic vortex beams the intrinsic OAM  is uniquely defined: it coincides with the total OAM, $L_z = {L}_z^{\rm int} = \ell$, and parallel translations of the $z$-axis keep it invariant \cite{Berry1998}. In contrast, parallel translations of the $z$ and $y$ axes, as well as  the definition of the centroid $X$, dramatically affect the OAM and its intrinsic part in STVPs. 
Due to the momentum $P_z \simeq k_0$, the OAM values are very sensitive to subwavelength $x \sim k_0^{-1}$ shifts. In particular, one can define the separation of the intrinsic and extrinsic parts of the OAM using the energy centroid ${\bf R}_W$ instead of ${\bf R}$ in Eq.~(\ref{eq_ext}). In this case, the STVP (\ref{eq_spectrum})
with $X_W =0$ will have the intrinsic OAM ${L}_y^{{\rm int}\prime} = \gamma \ell /2$. Thus, the definitions of the intrinsic and extrinsic OAM is a matter of convention for the pulse centroid. The advantage of the definitions (\ref{eq_R}) and (\ref{eq_ext}) is that the intrinsic OAM takes a universal form (\ref{eq_Ly_ext}) valid for both spatial and spatiotemporal vortices and arbitrary orientation of the OAM with respect to the propagation direction \cite{Bliokh2012,Wang2021,Zhang2022} (and also for different wave equations, see below).

Third, one can conclude from the above remarks that it is {\it impossible} to construct an STVP symmetric with respect to the $z$-axis in the $(z,x)$ plane. Indeed, the probability and energy centroids are inevitably shifted with respect to each other, and at least one of them is shifted with respect to the vortex. Moreover, the very concept of a unique phase singularity, common for monochromatic fields, becomes somewhat {\it uncertain} for spatiotemporal vortices. For instance, the charge-$\ell$ electric-field singularity in the STVP (\ref{eq_spectrum}) splits into a constellation of $\ell$ mutually-shifted charge-1 singularities in the vector-potential with the Fourier spectrum ${\bf A} = -i \omega^{-1} {\bf E}$ (in the Coulomb gauge), Fig.~\ref{Fig2}. The real-space distribution of this vector-potential and its centroid also differ from their electric-field counterparts. Thus, many basic features of spatiotemporal vortices become dependent on the choice of the wavefield (wavefunction), and one should be careful in applying concepts of monochromatic singular optics (where the vector-potential differs from the electric field by a constant factor) \cite{Soskin2001,Dennis2009} to spatiotemporal structured waves.

%Importantly, the intrinsic part of the OAM is robust to small perturbations of the pulse, while the extrinsic and total OAM can be very sensitive. For instance, if we construct a STVP with an additional amplitude factor $\sqrt{\omega}$ in the spectrum, i.e., $E_y \propto \sqrt{\omega} \exp(i\ell\phi)$, all the interfering plane waves will carry the same density of photons. This will result in $\bar{X} = \bar{L}_y^{\rm ext} =0$ and $\bar{L}_y = \bar{L}_y^{\rm int}$ given by Eq.~(\ref{eq_Ly_ext}). Such paraxial pulse looks very similar to the regular STVP, only with slight asymmetry of the energy-density distribution and shift of the {\it energy} centroid, Fig.~\ref{Fig1}(c). However, its total OAM differs dramatically from Eq.~(\ref{eq_Ly_ext}), and only the intrinsic OAM remains the same.

%%%%%%%%%%%%%%%%%%%%%%%%%%%%%%
\section{Acoustic STVPs}
%%%%%%%%%%%%%%%%%%%%%%%%%%%%%%
We now examine another example of STVPs, constructed for sound waves in a fluid or gas. These waves are described by the real-valued scalar pressure ${\cal P}({\bf r},t)$ and vector velocity ${\mathbfcal V}({\bf r},t)$ fields. Setting the medium parameters (the mass density, compressibility, and speed of sound) equal to 1, the complex plane-wave Fourier amplitudes of the fields, ${P}({\bf k}) e^{-i\omega({\bf k}) t}$ and ${\bf V}({\bf k}) e^{-i\omega({\bf k}) t}$, satisfy ${\bf V} = \omega^{-1}{\bf k}\, P$ with $\omega({\bf k}) = k$. Similarly to Eqs.~(\ref{eq_W})--(\ref{eq_J}), the energy, momentum, and OAM of a paraxial sound wavepacket can be written as \cite{LLfluid,Bliokh2019_II}: 
\begin{align}
\label{eq_W_sound}
& W = \frac{1}{2N}\! \int ({\mathcal{P}}^2 + {\mathbfcal{V}}^2)\, d^3{\bf r} 
= \frac{\int |P|^2\, d^3{\bf k}}{\int  \omega^{-1}|P|^2\, d^3{\bf k}} \simeq \omega_0, \\ 
\label{eq_P_sound}
& {\bf P} = \frac{1}{N}\! \int {\mathcal{P}}  {\mathbfcal{V}}\, d^3{\bf r} 
= \frac{\int \omega^{-1} {\bf k}\, |P|^2\, d^3{\bf k}}{\int  \omega^{-1}|P|^2\, d^3{\bf k}} \simeq {\bf k}_0 \,, \\
\label{eq_L_sound}
& {\bf L} =  \frac{1}{N}\! \int \! {\bf r} \times\! ({\mathcal{P}}  {\mathbfcal{V}})\, d^3{\bf r} 
= \frac{\int\!  \omega^{-1} \!{P}^* (-i{\bf k} \times \! {\bm \nabla}_{\bf k}) {P} \, d^3{\bf k}}{\int  \omega^{-1}|P|^2\, d^3{\bf k}},
\end{align}
where $N \propto \int  \omega^{-1}|P|^2\, d^3{\bf k}$ is the number of phonons in the pulse.

Equations (\ref{eq_W_sound})--(\ref{eq_L_sound}) show that in terms of the pressure-field `wavefunction' $P({\bf k})$ the expectation values for {\it phonons} acquire the same $ \omega^{-1}$ factor as in the case of photons. Therefore, the spatiotemporal OAM calculations (\ref{eq_spectrum})--(\ref{eq_Ly_ext}) remain the same. Although in this representation the definitions for spin-0 phonons and spin-1 photons coincide, the equations of motion for sound waves are actually mapped onto the massless Klein-Gordon equation with the wavefunction $\psi$: ${\cal P} = - \partial_t \psi$ \cite{Bliokh2019PRL}. Constructing the STVP in terms of plane waves $\psi$ with equal amplitudes yields different values of the extrinsic and total OAM (see below); the difference between the `wavefunctions' ${\cal P}$ and $\psi$ is similar to the difference between the electric field and vector potential in the electromagnetic case.

%%%%%%%%%%%%%%%%%%%%%%%%%%%%%%
\section{Klein-Gordon STVPs}
%%%%%%%%%%%%%%%%%%%%%%%%%%%%%%
Let us now consider STVPs in the Klein-Gordon wave equation which describes both massless and massive spin-0 quantum particles. Using the complex wavefunction $\psi({\bf r},t)$ and its Fourier transform $\tilde\psi({\bf k})$, the energy, momentum, and angular momentum of a paraxial Klein-Gordon wavepacket read \cite{BLP,Bliokh2012PRL}:
\begin{align}
\label{eq_W_KG}
& W =  \frac{1}{N} \int w\, d^3{\bf r}
= \frac{\int \omega^2 |\tilde{\psi} |^2\, d^3{\bf k}}{\int \omega |\tilde{\psi} |^2\, d^3{\bf k}} 
\simeq \omega_0\,, \\ 
\label{eq_P_KG}
& {\bf P} = \frac{1}{N} \int {\bf p}\, d^3{\bf r} 
= \frac{\int \omega {\bf k} |\tilde{\psi}|^2\, d^3{\bf k}}{\int \omega |\tilde{\psi} |^2\, d^3{\bf k}} 
\simeq {\bf k}_0 , \\
& {\bf L} = \frac{1}{N} \int {\bf r}\times{\bf p}\, d^3{\bf r}
=  \frac{\int  {\omega}\, \tilde{\psi}^* (-i{\bf k} \times {\bm \nabla}_{\bf k}) \tilde{\psi} \, d^3{\bf k}}{\int \omega |\tilde{\psi} |^2\, d^3{\bf k}}\,.
\label{eq_L_KG} 
\end{align}
Here $w= (|\partial_t \psi |^2 + |{\bm \nabla} \psi |^2 + m^2 |\psi |^2)/2$ and ${\bf p} = -{\rm Re}\! \left[(\partial_t \psi^*){\bm \nabla} \psi \right]$ are the energy and momentum densities for the Klein-Gordon particles, $N = \int \rho \, d^3{\bf r} \propto \int \omega |\tilde{\psi} |^2\, d^3{\bf k}$ is the number of particles, $\rho = {\rm Im} \!\left[(\partial_t \psi^*)\psi \right]$ is the probability density, $m$ is the mass, we use the units $c=\hbar = 1$, and the dispersion relation is $\omega^2 = k^2 + m^2$. 

Thus, the definitions of the Klein-Gordon wavefunction and expectation values involves the factor $\omega$ instead of $\omega^{-1}$ in the photon case. This is because we deal with spin-0 particles. (For spin-1/2 Dirac particles there is no factor.)
Akin to the photon case, assuming the Bessel-type STVP spectrum (\ref{eq_spectrum}) with $\tilde{\psi} \propto \exp(i\ell\phi)$, we calculate the $x$-coordinate of the particle probability centroid:
\begin{align}
\label{eq_X_KG}
{X} = \frac{\int x \rho \, d^3{\bf r}}{\int \rho \, d^3{\bf r}}
= \frac{\int  {\omega}\, \tilde{\psi}^* (i \partial_{k_x}) \tilde{\psi} \, d^3{\bf k}}{\int \omega |\tilde{\psi} |^2\, d^3{\bf k}} 
\simeq - \frac{\ell\, k_0}{2\gamma\, \omega_0^2}\,,
\end{align}
where $\omega_0 = \sqrt{k_0^2 +m^2}$. This shift is directed oppositely to Eq.~(\ref{eq_X}). Calculating the OAM (\ref{eq_L_KG}), and using Eqs.~(\ref{eq_ext}) and (\ref{eq_X}), we find that the total, intrinsic, and extrinsic OAM of the Klein-Gordon STVP:
\begin{equation}
\label{eq_Ly_KG}
{L}_y = {L}_y^{\rm int} + {L}_y^{\rm ext} \simeq  \ell\, \frac{\gamma + \gamma^{-1}}{2} + \ell\, \frac{ k_0^2}{2\gamma \omega_0^2} \,.
%= \ell\, \frac{\gamma + 2 \gamma^{-1}}{2} \,.
\end{equation}
Notably, the intrinsic OAM is the same as in the photon case (\ref{eq_Ly_ext}).
We checked the analytical results (\ref{eq_X_KG}) and (\ref{eq_Ly_KG}) by numerical calculations using the wavefunction $\psi({\bf r},t)$ and real-space integrals in Eqs.~(\ref{eq_L_KG}) and (\ref{eq_X_KG}). To construct a Klein-Gordon STVP with ${X} = {L}_y^{\rm ext} = 0$ and ${L}_y = {L}_y^{\rm int}$, one has to use the vortex wavefunction $\tilde{\psi} \propto {\omega}^{-1/2}\exp(i\ell\phi)$, which involves plane waves with equal particle densities. 

Note that the Klein-Gordon STVP with a symmetric distribution $|\psi ({\bf r})|^2$ has an asymmetric energy-density distribution $w ({\bf r})$ with the centroid $X_W = 2 X$. Therefore, an alternative definition of the intrinsic OAM with respect to $X_W$ yields ${L}_y^{{\rm int}\prime} = \left[\gamma + \gamma^{-1}(1-k_0^2/\omega_0^2)\right]\!/{2}$.

In addition to all the properties described for electromagnetic STVPs, the Klein-Gordon case allows one to trace the OAM behaviour from the relativistic massless limit $k_0 = \omega_0$ to the nonrelativistic limit $k_0/\omega_0 \to 0$. In the latter case, ${X} =X_W= {L}_y^{\rm ext} =0$, and ${L}_y = {L}_y^{\rm int} = {L}_y^{{\rm int}\prime}$ is unambiguously given by Eq.~(\ref{eq_Ly_ext}) \cite{Bliokh2012,Bliokh2021}. Thus, all the peculiarities related to the centroid shift and extrinsic OAM have {\it relativistic} origin. 

Indeed, the notion of the `mass centroid' of an extended body is well-defined in non-relativistic mechanics and it unambiguously determines the separation of the intrinsic and extrinsic OAM \cite{LL_mech}. In relativistic mechanics, the centroid of the body becomes frame-dependent and can be defined in different ways \cite{LLfield,Pryce1948,Bliokh2012PRL}. Notably, using the seminal approach by Pryce \cite{Pryce1948}, the expectation value of the quantum-mechanical operator of the relativistic mass/energy centroid corresponds to the probability centroid ${\bf R}$ rather than the energy centroid ${\bf R}_W$ \cite{Bliokh2017PRA}. This is because the expectation value of some combination of operators differs from the analogous combination of the expectation values of the operators.

%%%%%%%%%%%%%%%%%%%%%%%%%%%%%%
\section{Generalization to arbitrary spin}
%%%%%%%%%%%%%%%%%%%%%%%%%%%%%%
It is instructive to generalize the photon and Klein-Gordon results to STVPs in a relativistic wave equation with arbitrary spin $s$ and mass. Assuming the momentum-representation wave function $\tilde{\psi} \propto \omega^{n/2} \exp(i\ell\phi)$, and using the density of particles in a plane wave $\rho \propto \omega^{(1-2s)} |\tilde{\psi}|^2$, we find the total OAM and positions of the particle and energy centroids:
\begin{align}
\label{eq_Ly_general}
& {L}_y \simeq   \ell\, \frac{\gamma + \gamma^{-1}}{2} + \ell\, (1-2s+n) \frac{ k_0^2}{2\gamma \omega_0^2}  \,. \\
\label{eq_X_general}
& {X} \simeq - \ell \, (1-2s+n)\frac{k_0}{2\gamma \omega_0^2}\,,~~
{X}_W \simeq {X}  - \ell \frac{k_0}{2\gamma \omega_0^2}\,.
%= \ell\, \frac{\gamma + 2 \gamma^{-1}}{2} \,.
\end{align}
For example, the electromagnetic STVPs in Figs.~\ref{Fig1}(b) and (c) correspond to $k_0=\omega_0$, $s=1$, $n=0$ and $n=1$, respectively.
 
Remarkably, there is only one combination in Eqs.~(\ref{eq_Ly_general}) and (\ref{eq_X_general}) that acquires a universal form (\ref{eq_Ly_ext}) \cite{Bliokh2012,Bliokh2021}, independent of spin, mass, etc., and similar to the OAM of monochromatic beams: 
\begin{align}
\label{eq_Ly_int_general}
{L}_y^{\rm int} \simeq {L}_y +k_0 {X} \simeq \ell \, \frac{\gamma + \gamma^{-1}}{2}  \,. 
\end{align}
That is why we advocate the definition (\ref{eq_ext}) of the intrinsic OAM. Alternatively, defining the intrinsic OAM with respect to the energy centroid yields 
\begin{align}
\label{eq_Ly'_int_general}
{L}_y^{{\rm int} \prime} \simeq {L}_y +k_0 {X}_W \simeq \ell \, \frac{\gamma + \gamma^{-1}(1-k_0^2/\omega_0^2)}{2}  \,. 
\end{align}
In the massless case, $k_0 =\omega_0$, this coincides with the OAM (\ref{eq_Ly}) advocated in \cite{Hancock2021}. In the nonrelativistic limit, both definitions converge to the same universal form. 

%%%%%%%%%%%%%%%%%%%%%%%%%%%%%%
\section{Conclusions}
%%%%%%%%%%%%%%%%%%%%%%%%%%%%%%
We have resolved a recent controversy in calculations of the OAM of electromagnetic STVPs \cite{Bliokh2012,Chong2020,Bliokh2021,Hancock2021}, and found that the OAM properties of paraxial STVPs are extremely sensitive to a number of factors: definition of the pulse centroid, spin of the particle, its mean momentum (relativistic or non-relativistic), and choice of the wavefunction. This is in sharp contrast to monochromatic paraxial vortex beams, where the intrinsic OAM is unambiguous and independent of all these factors \cite{Allen1992, Allen_book, Bekshaev_book, Andrews_book, Molina2007, Franke2008, Bliokh2015PR, Hefner1999, Lekner2006, Volke2008, Demore2012, Anhauser2012, Marzo2018,Bliokh2019_II, Bliokh2007, Uchida2010, Verbeeck2010, McMorran2011, Bliokh2017, Lloyd2017, Larocque2018, Clark2015, Luski2021,Berry1998}. In the spatiotemporal case, globally-symmetric vortices with a unique center are simply impossible: a relativistic object with an intrinsic angular momentum is inevitably asymmetric and delocalized. 
Nonetheless, we found that calculating the intrinsic OAM with respect to the probability centroid of the particle results in the clear universal expression \cite{Bliokh2012,Bliokh2021} independent of all the above subtleties and similar to the monochromatic vortex case.

\blue{Furthermore, we have described the OAM properties of STVPs in other wave systems: acoustic and quantum-relativistic with different spins and masses. Using the suggested approach based on the probability centroid, the transverse intrinsic OAM takes the same universal form in all of these cases. It is worth noticing that the OAM of acoustic pulses is determined by the {\it mechanical} angular momentum of the medium particles moving due to the wave-induced Stokes drift \cite{Bliokh2022,Bliokh2022PRA}. This makes it directly observable for classical waves in fluids or gases \cite{Francois2017,Bliokh2022}.}

We hope that this work sheds light on the intricate analysis of the OAM and other properties of spatiotemporal vortices. 
%Our results span a wide range of wave systems: optical, acoustic, and quantum-relativistic. 
For further progress in this problem, it is important to examine possible methods of the OAM measurements in STVPs of various nature. Atomic transitions in interactions with optical pulses could be one of such possibilities \cite{Enk1994_II,Afanasev2018}. \blue{Also, the radiation torque on large absorbing  particles (larger than the pulse size) could serve as a measure of the OAM in optical \cite{He1995, Simpson1997} and acoustic \cite{Volke2008, Demore2012, Anhauser2012} STVPs. In these cases, an accurate analysis of the wave-matter interactions involving non-monochromatic pulses is necessary.} 

%\vspace*{0.3cm}
\begin{acknowledgments}
%%%%%%%%%%%%%%%%%%%%%%%%%%%%%%
%\section*{ACKNOWLEDGMENTS}
%%%%%%%%%%%%%%%%%%%%%%%%%%%%%%
I acknowledge fruitful debates with Howard M. Milchberg, which stimulated this work, and helpful discussions with Miguel A. Alonso.
\end{acknowledgments}

%\newpage

%\pagebreak

\bibliography{References_STVP}

\end{document}